# Approaching quantum anomalous Hall effect in proximity-coupled YIG/graphene/h-BN sandwich structure

Chi Tang[1], Bin Cheng[1], Mohammed Aldosary[1], Zhiyong Wang[1], Zilong Jiang[1], K. Watanabe[3], T. Taniguchi[3], Marc Bockrath[1,2], and Jing Shi[1,a]

[1]Department of Physics and Astronomy, University of California, Riverside, CA 92521, USA

[2]Department of Physics, The Ohio State University, Columbus, OH 43210, USA

[3]Advanced Materials Laboratory, National Institute for Materials Science, Tsukuba, Ibaraki 305-0044, Japan

Quantum anomalous Hall state is expected to emerge in Dirac electron systems such as graphene under both sufficiently strong exchange and spin-orbit interactions. In pristine graphene, neither interaction exists; however, both interactions can be acquired by coupling graphene to a magnetic insulator (MI) as revealed by the anomalous Hall effect. Here, we show enhanced magnetic proximity coupling by sandwiching graphene between a ferrimagnetic insulator yttrium iron garnet (YIG) and hexagonal-boron nitride (h-BN) which also serves as a top gate dielectric. By sweeping the top-gate voltage, we observe Fermi level-dependent anomalous Hall conductance. As the Dirac point is approached from both electron and hole sides, the anomalous Hall conductance reaches ¼ of the quantum anomalous Hall conductance $2e^2/h$. The exchange coupling strength is determined to be as high as 27 meV from the transition temperature of the induced magnetic phase. YIG/graphene/h-BN is an excellent heterostructure for demonstrating proximity-induced interactions in two-dimensional electron systems.

Long-range ferromagnetic order in two-dimensional electron systems (2DES) has long been sought to involve the spin degree of freedom in spatially confined quantum systems[1-5]; however, the transition metal doping approach failed to deliver high-temperature ferromagnetism. Recently it became possible with the advent of 2D layered materials such as graphene[6,7] and other van der Waals (vdW) materials[8-11]. In the former, ferromagnetism is introduced by proximity coupling in a heterostructure comprising of graphene and magnetic insulator; while in the later, the material itself is a spontaneously ferromagnetically ordered 2D vdW crystal, either a 2D Heisenberg or an Ising ferromagnet. Such 2D ferromagnetic systems provide unprecedented opportunities to create novel heterostructures for exploiting spin-dependent phenomena in spatially confined electron systems.

A particularly interesting quantum phenomenon in graphene was theoretically proposed by Qiao *et al.*[12,13], namely the quantum anomalous Hall effect (QAHE), which emerges in the presence of finite exchange interaction and spin-orbit coupling (SOC). Under these two interactions, a topological band gap or an exchange gap is opened up at the Dirac point which consequently gives rise to a quantized anomalous Hall conductance of $\pm 2e^2/h$. However, neither interaction is present in pristine graphene. Ferromagnetic order has recently been demonstrated via proximity coupling in both yttrium iron garnet(YIG)/graphene[6] and EuS/graphene[7], where YIG or EuS breaks the time reversal and inversion symmetries and therefore serves as the source of these interactions. To ultimately realize the QAHE in graphene, besides sufficiently strong exchange and SOC, graphene needs to have relatively weak disorder so that the energy scale associated with the disorder is smaller than the exchange gap. Therefore, a main challenge is to maximize the exchange gap and simultaneously minimize the disorder energy scale. Here, we show an enhanced proximity exchange effect in a YIG/graphene/h-BN sandwich structure revealed by the anomalous Hall effect (AHE). In this study, the h-BN layer replaces the poly-methyl methacrylate (PMMA) layer used in previous devices[6], which not only makes an efficient top gate but also protects graphene underneath to preserve its high mobility[14]. As a result, the AHE conductance shows a clear gate voltage dependence as the Fermi level is tuned towards the Dirac point, which is expected from a gapped Dirac spectrum. Moreover, the AHE conductance reaches ¼ of the QAHE conductance $2e^2/h$. The induced magnetic phase transition temperature in the heterostructure is as high as room temperature, with an exchange coupling strength of ~27 meV.

We first epitaxially grow ~ 20 nm thick YIG films using pulsed laser deposition on (111)-oriented gadolinium gallium garnet substrate[15]. The surface topography of the YIG films is characterized by atomic force microscope (AFM) with a root-mean-square (rms) roughness of ~1 - 2 Å over a 2 × 2 µm$^2$ scan area, as shown in Fig. 1(b). Due to the interfacial nature of the proximity coupling, the atomically flat surface of YIG films is critical to create a strong exchange interaction. Standard Hall bar patterned graphene devices are fabricated on 290 nm-thick Si/SiO$_2$ substrates. Single-layer graphene flakes are located under an optical microscope and confirmed by Raman spectroscopy. We use a 20 – 30 nm thick h-BN layer via a micromechanical transfer process[16] to cover the entire graphene region, which simultaneously protects graphene from chemical solvent or resist contaminations and serves as an effective top gate dielectric layer. After top gate electrodes are fabricated and tested, we transfer the entire functional graphene/h-BN devices from Si/SiO$_2$ substrates to YIG utilizing a previously developed transfer technique[6,17]. The optical images of graphene/h-BN devices before and after transfer are illustrated in Fig. 1(c), which show a successful transfer. Transport measurements are performed on the same graphene/h-BN devices before and after transfer to track the magneto-transport responses on different substrates.

The soft magnetic hysteresis of a bare YIG film with the magnetic field applied along the out-of-plane direction shown in Fig. 2 (a) confirms the in-plane magnetic anisotropy arising primarily from the shape anisotropy[15]. Additional growth- or interface strain-induced in-plane magnetic anisotropy contributes to a higher saturation magnetic field (usually between 2000 and 3000 Oe) than the demagnetizing field $4\pi M_s$[18-20]. When graphene is placed on SiO$_2$, the Hall resistance is linear in external magnetic field, which originates from the ordinary Hall effect (OHE). After the linear background is subtracted, no net signal is left as shown in Fig. 2 (a). In the same device transferred to YIG, a clear nonlinear Hall curve stands out after subtraction of a linear OHE background also shown in Fig. 2 (a). Below the Hall curves, the magnetic hysteresis of the YIG film measured with the out-of-plane magnetic fields is displayed. It is clear that the nonlinear Hall curve resembles the magnetization of the underlying YIG layer. The observed nonlinear Hall behavior can arise from the following possible mechanisms: (i) the AHE response of graphene in which the carriers are spin polarized due to the interfacial exchange coupling with YIG; (ii) the Lorentz force induced nonlinear Hall effect either from the coexistence of two types of carriers in graphene[21] or the stray magnetic field produced by YIG. As will be discussed, the gate dependence measurements exclude the latter possibility, which favors the first scenario, i.e. induced ferromagnetism in graphene due to the exchange coupling with the YIG layer. Here we ascribe the nonlinear Hall response to the AHE.

To quantitatively characterize the exchange coupling strength in YIG/graphene/h-BN, the AHE response is studied over a wide range of temperatures from 13 K to 300 K. Fig. 2(b) shows the data taken at several selected temperatures with a top gate voltage of 0.9 V. The AHE resistance progressively decreases as the temperature increases, but persists up to 300 K. Note that the AHE resistance is proportional to the spontaneous magnetization and also has power-law dependence on the longitudinal resistance[22]. Since the latter is a smooth function of temperature, near the ferromagnetic transition temperature $T_c$, the AHE resistance can be expressed as $R_{AH} \sim (T - T_c)^\beta$ with the critical component $\beta = 0.5$. By fitting the temperature dependence data, we extract the transition temperature $T_c \sim 308$ K, or $k_B T_c \sim 27$ meV in exchange energy, which is higher than what was previously reported[6].

As mentioned earlier, a trivial cause of the nonlinear Hall signal in YIG/graphene/h-BN is the Lorentz force, either from the coexistence of two types of carriers or from the stray magnetic field produced by the YIG substrate. Such a nonlinear Hall response should reverse its sign when the carriers change from electron- to hole-type or vice versa, whereas the AHE sign does not have to change as the carrier type switches. Thus, we measure the nonlinear Hall response at each fixed gate voltage and repeat for a range of top gate voltages which set different carrier densities on both hole and electron sides. Fig. 3(a) shows the sheet resistance of YIG/graphene/h-BN as a function of the top gate voltage $V_{tg}$ measured at 13 K. The Dirac point $V_{DP}$ of the graphene device is at 0.5 V, very close to zero. The Hall mobility is as high as 18,400 cm$^2$/Vs, about the same order of magnitude as in SiO$_2$/graphene/h-BN, indicating no negative effect on carrier mobility due to the transfer process. The mobility is at least 3 times as high as that in graphene on SiO$_2$ without h-BN. Due to the thinner h-BN sheet, only much smaller applied gate voltages (decreased by a factor of 15) are needed to tune the carrier density over a wide range.

To avoid the region where electrons and holes coexist in the vicinity of the Dirac point which can also produce a strong and complex nonlinear OHE, we measure the Hall voltages in YIG/graphene/h-BN only with $V_{tg} \leq 0\ V$ and $V_{tg} \geq 0.9\ V$, as indicated in Fig. 3(b). The sign of the anomalous Hall resistance is independent of the carrier type. If the Hall response were generated by the Lorentz force, the sign of nonlinear Hall resistance would switch once the carrier type changes. The observed same nonlinear Hall sign clearly excludes the Lorentz force related mechanism, and therefore unambiguously demonstrates the anomalous Hall origin due to the SOC in the proximity-induced ferromagnetic phase of graphene. As expected for QAHE insulators[12,13,22,23], the intrinsic AHE from the Berry curvature indeed has the same sign on both electron and hole sides in unquantized regions where the Fermi level is outside the exchange gap.

We calculate the anomalous Hal conductance $\sigma_{AH}$ as a function of the top gate voltage and plot it in Fig. 4. At large gate voltages on both sides, $\sigma_{AH}$ is relatively small. As the gate voltage decreases to approach the Dirac point from both sides, $\sigma_{AH}$ gradually increases and follows the trend predicted by *Qiao et al.*[12,13] Here we intentionally stay away from the two-carrier dominated region near the Dirac point. Over the measured gate voltage range, $\sigma_{AH}$ is clearly not quantized. The maximum $\sigma_{AH}$ in our best device reaches ¼ of the quantum anomalous Hall conductance, $2e^2/h$. Both the magnitude and the clear gate voltage dependence of $\sigma_{AH}$ indicate much improved proximity-induced exchange and SOC strengths compared to the previous YIG/graphene/PMMA devices, thanks to the h-BN that preserves the high quality of graphene sheet and reduces the disorder strength (~ 13 meV). However, the absence of the QAHE plateau suggests a small exchange gap which is smeared out by thermal fluctuations and disorder. The exchange gap is determined by the small interaction of the exchange and SOC. Since we have achieved relatively large exchange interaction, the small exchange gap is mainly due to relatively weak SOC. To ultimately demonstrate the QAHE in graphene, greatly enhanced SOC is clearly required. Recent experiments indicate that it is possible to drastically enhance the Rashba SOC via proximity coupling with transition metal dichalcogenide materials such as $WSe_2$[24,25].

In summary, the observed AHE demonstrates the existence of long-range ferromagnetic order in graphene proximity coupled with a magnetic insulator. The maximum anomalous Hall conductance reaches ¼ of the QAHE conductance $2e^2/h$. The exchange coupling strength is determined to be as high as 27 meV, indicating an effective role the h-BN played in promoting both exchange and SOC and reducing the disorder strength. Further exploration of incorporating a transition metal dichalcogenide layer into the sandwich structure to enlarge SOC in graphene and utilizing thulium iron garnet with robust perpendicular magnetic anisotropy[26] is promising to realize the QAHE at high temperatures and zero magnetic field.

YIG film growth and characterizations, YIG/graphene heterostructure device fabrication, transport measurements, and data analysis were supported in part by DOE BES Award No. DE-FG02-07ER46351. Construction of the transfer microscope and device characterizations were supported by NSF-ECCS under Awards No. 1202559 and NSF-ECCS and No. 1610447. Exfoliation and transfer of h-BN were supported by DOE ER 46940-DE-SC0010597.

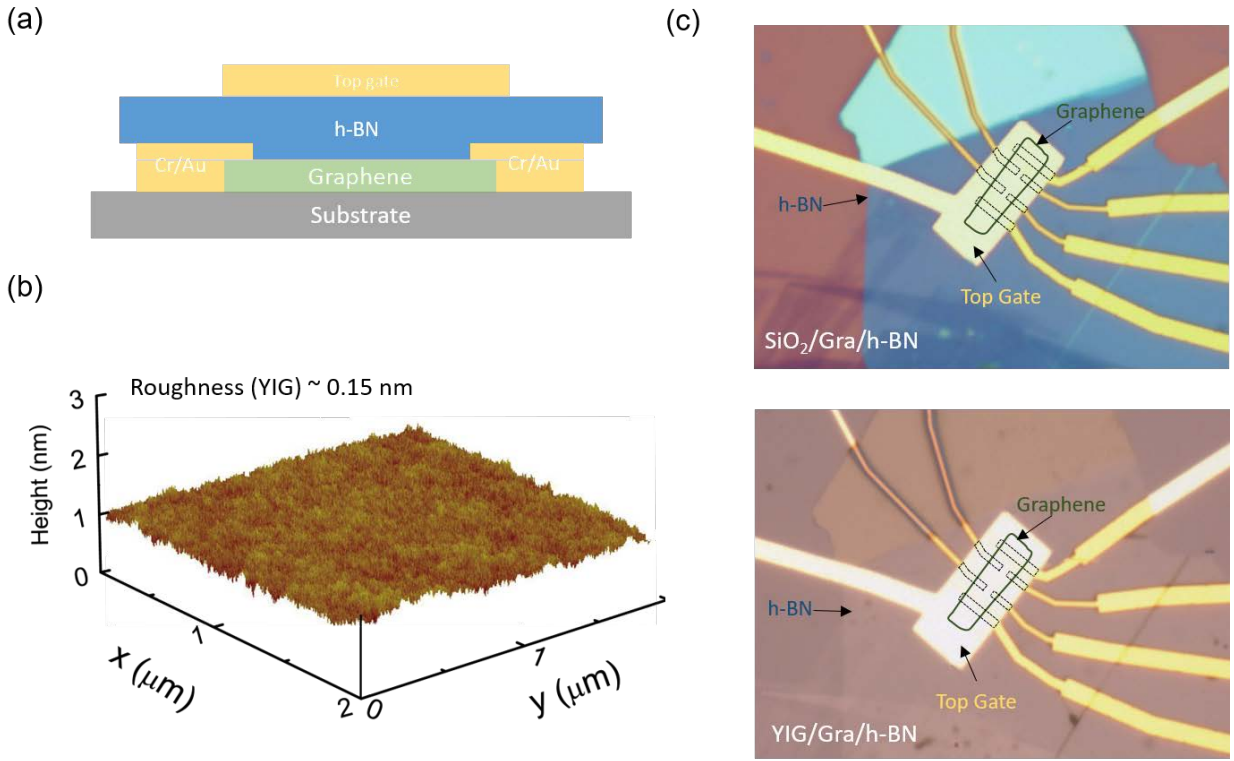

Fig. 1. (a) Schematic view of YIG/graphene/h-BN device; (b) AFM image of a typical YIG film grown by pulsed laser deposition with roughness ~ 0.15 nm across a 2 × 2 μm$^2$ scan area; (c) Top: exfoliated graphene on SiO$_2$ covered with h-BN. Bottom: transferred graphene/h-BN devices on YIG substrate.

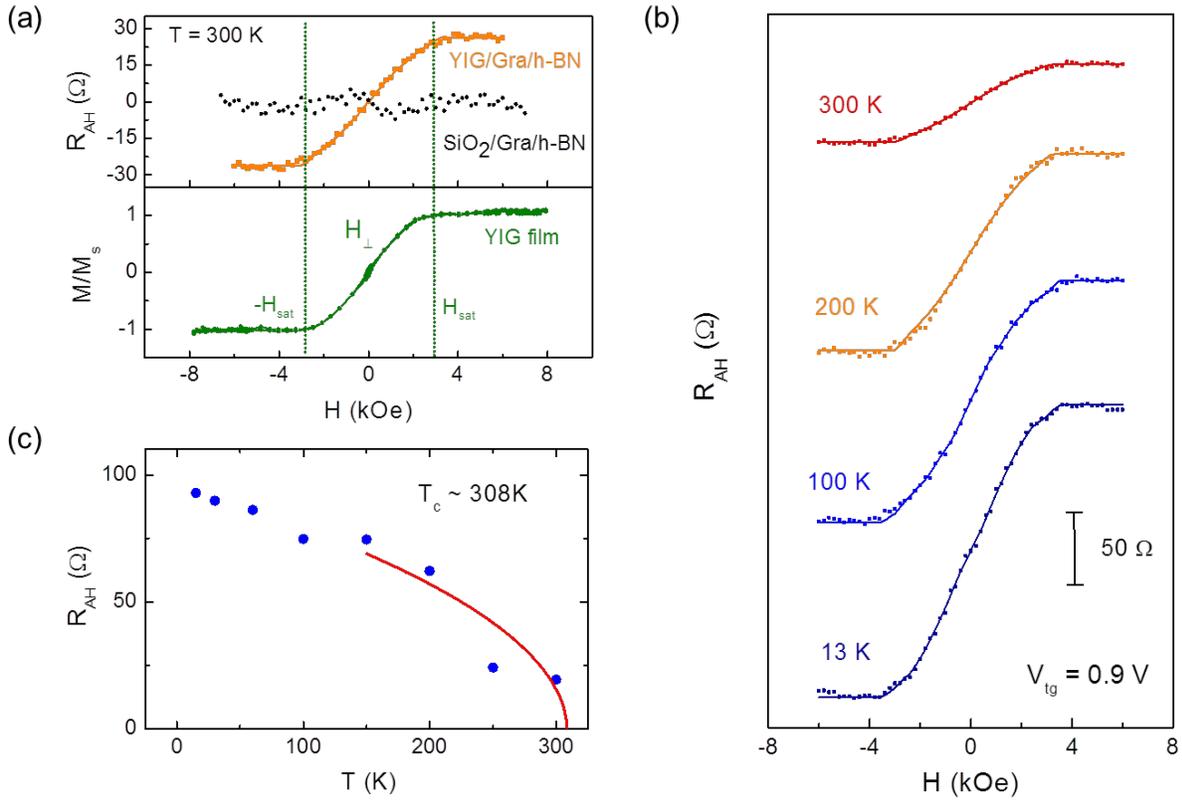

Fig. 2. (a) Top: Anomalous Hall resistance of graphene/h-BN on YIG (orange) and $SiO_2$ (black) at 300 K after the OHE background is subtracted. No nonlinear response is left in graphene on $SiO_2$ whereas there is a clear anomalous Hall signal in graphene transferred onto YIG. Bottom: Magnetic hysteresis behavior of YIG film (green) when an external magnetic field is applied perpendicular to graphene. The nonlinear Hall resistance in YIG/graphene/h-BN follows the magnetization of the YIG film. (b) Anomalous Hall resistance curves of YIG/graphene/h-BN at the top gate voltage of 0.9 V measured from 13 to 300 K; (c) Temperature dependence of anomalous Hall resistance of YIG/graphene/h-BN. The magnetic phase transition temperature $T_c$ of the induced ferromagnetism in graphene is extracted to be 308 K by using $R_{AH} \sim (T - T_c)^{\beta}$ with the critical component $\beta = 0.5$.

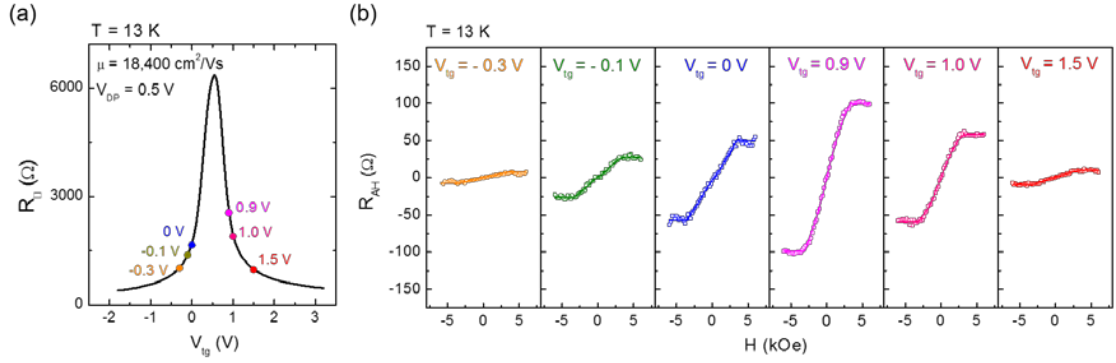

Fig. 3. (a) Top gate voltage dependence of the sheet resistance of graphene sandwiched between YIG and h-BN measured at 13 K with a mobility of 18,400 cm$^2$/Vs and the Dirac point near 0.5 V. Several top gate voltages are selected to show the anomalous Hall resistance at different carrier densities on both electron and hole dominated regions. (b) Anomalous Hall resistance in YIG/graphene/h-BN at different top gate voltages. The anomalous Hall resistance sign remains the same for both electron and hole carrier types.

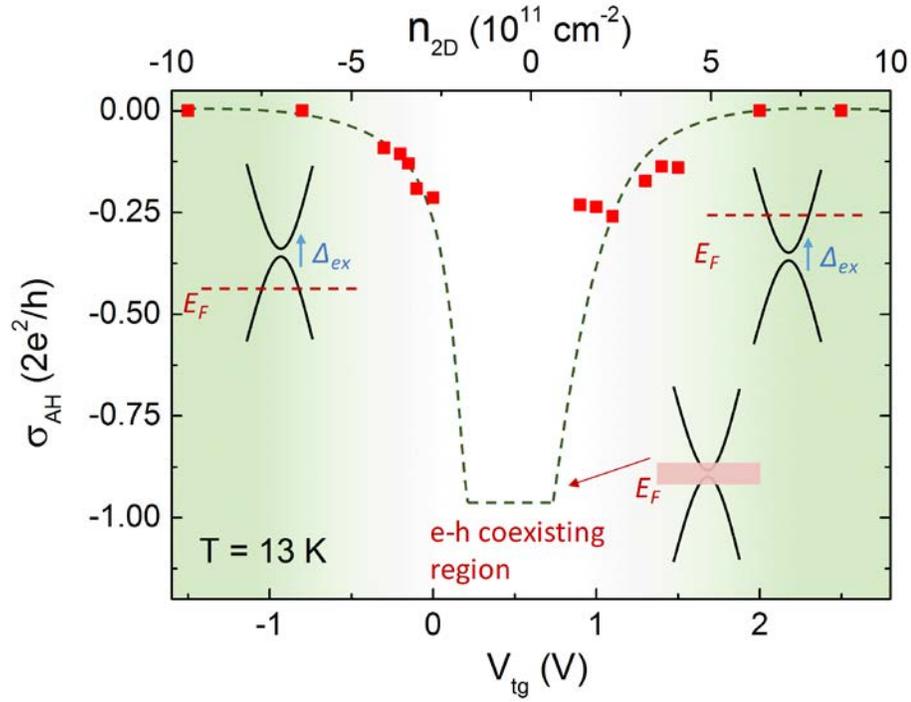

Fig. 4. Top gate voltage and carrier density dependence of anomalous Hall conductance measured at 13 K. Red squares are experimental data and dashed green curve is drawn for the purpose of eye guidance. The green area marks the electron or hole-dominated region where clear AHE is observed. The largest anomalous Hall conductance in YIG/graphene/h-BN reaches ¼ of the quantum anomalous Hall conductance. The white region is the e-h coexisting region where the Fermi level is too close to the Dirac point and additional oscillatory features are observed due to multi-carriers in this region.